\documentstyle[twocolumn,aps,epsf]{revtex}
\begin{document}
\draft
\twocolumn[\hsize\textwidth\columnwidth\hsize\csname
@twocolumnfalse\endcsname


\title{Spin Fluctuation Induced Superconductivity
Controlled by Orbital Fluctuation}

\author{T Takimoto\dag, T Hotta\dag, T Maehira\dag,
and K Ueda\ddag\dag}

\address{\dag\ Advanced Science Research Center,
Japan Atomic Energy Research Institute, Tokai, Ibaraki 319-1195, Japan}
\address{\ddag\ Institute for Solid State Physics, University of Tokyo,
5-1-5 Kashiwa-no-ha, Kashiwa, Chiba 277-8581, Japan}

\date{\today}
\maketitle

\begin{abstract}
A microscopic Hamiltonian reflecting the correct symmetry of 
$f$-orbitals
is proposed to discuss superconductivity in heavy fermion systems.
In the orbitally degenerate region in which not only spin fluctuations 
but also orbital fluctuations develop considerably,
cancellation between spin and orbital fluctuations destabilizes
$d_{x^{2}-y^{2}}$-wave superconductivity. 
Entering the non-degenerate region by increasing
the crystalline electric field, 
$d_{x^{2}-y^{2}}$-wave superconductivity
mediated by antiferromagnetic spin fluctuations emerges 
out of the suppression of orbital fluctuations.
We argue that the present scenario can be applied to
recently discovered superconductors CeTIn$_{5}$
(T=Ir, Rh, and Co).
\end{abstract}


\vskip2pc]
\narrowtext

Unconventional superconductivity has been one of central issues
in the research field of strongly correlated electron systems.
Especially since the discovery of high temperature superconductivity
in cuprates,
much effort has been focussed on elucidating the mechanism of
unconventional superconductivity, clarifying that a crucial role
is played by antiferromagnetic (AF)
``spin fluctuations'' (SF) \cite{Moriya}.
The importance of AFSF is widely recognized, for instance,
in $4f$- and $5f$-electron superconducting materials such as
${\rm CeCu_2Si_2}$ \cite{Steglich} and ${\rm UPd_2Al_3}$ \cite{Geibel} 
as well as in organic superconductors such as $\kappa$-(BEDT-TTF)
\cite{organic}.
Thus, it is widely believed that a broad class of
unconventional superconductors originates from SF.

In $d$- and $f$-electron systems, however, 
the potential importance of {\it orbital} degrees of freedom
has recently been discussed intensively.
In fact, ``orbital ordering'' is found to be a key issue for
understanding microscopic aspects of the charge-ordered phase
in colossal magnetoresistive manganites \cite{Dagotto}.
This orbital ordering is primarily relevant to 
the insulating phase,
while ``orbital fluctuations'' (OF) should be significant
in the metallic phase.
Recently, the effects of OF have attracted attention,
since it is hoped to provide a new scenario for superconductivity
\cite{Takimoto}.
Especially from a conceptual viewpoint,
it is important to clarify how superconductivity emerges
when both SF and OF play active roles.

As a typical material for investigating superconductivity
in a system with both SF and OF,
let us introduce the recently discovered heavy-fermion superconductors
${\rm CeTIn_5}$ (T=Ir, Rh, and Co) \cite{Ce115} with
the ${\rm HoCoGa_{5}}$-type tetragonal crystal structure.
Due to this structure and to strong correlation effects,
the similarity with cuprates has been emphasized.
In particular, AFSF also plays an essential role in the Ce-115 system,
since it exhibits quasi two-dimensional Fermi surfaces
\cite{Haga}
and the AF phase exists next to the superconducting state
in the phase diagram of ${\rm CeRh_{1-x}Ir_xIn_{5}}$
\cite{Pagliuso}.
In fact, a line node in the gap function has been observed
in CeTIn$_5$ by various experimental techniques \cite{gap}.
It may be true that superconductivity itself is due to AFSF,
but an important role for the OF has been overlooked
in spite of the fact that both SF and OF are originally included
in the ground-state multiplet of Ce$^{3+}$ ion.
In actual materials, superconductivity occurs
in a situation where OF is suppressed,
since orbital degeneracy is lifted by the effect of
the crystalline electric field (CEF).
Thus, we envision a scenario where OF controls
the stability of the superconductivity even though it originates 
from AFSF.

In this Letter, we investigate superconductivity based on the
orbitally degenerate Hubbard model constructed
by the tight-binding method \cite{Maehira}.
Solving the gap equation with the pairing interaction evaluated using
the random phase approximation (RPA),
we obtain several superconducting phases
around the spin and orbital ordered phases.
In the orbitally degenerate region 
where both SF and OF are developed,
it is found that singlet superconductivity is suppressed
due to the competition between them, while 
triplet superconductivity is favored since they are cooperative
in this case.
When a level splitting is included to lift the orbital degeneracy,
$d_{x^{2}-y^{2}}$-wave superconductivity due to AFSF is stabilized
in the vicinity of the AF phase.
Thus, we claim that AFSF-induced supercondutivity
in Ce-115 systems is substantiated in consequence of
the suppression of OF.



In order to construct the microscopic model for $f$-electron systems,
let us start our discussion from a local basis of 
the Ce$^{3+}$ ion.
Among the 14-fold degenerate $4f$-electron states,
due to the effect of strong spin-orbit coupling,
only the $j$=5/2 sextuplet effectively contributes to
the low-energy excitations ($j$ is total angular momentum).
This sextuplet is further split into a $\Gamma_7$ doublet
and a $\Gamma_8$ quadruplet due to the effect of cubic CEF, 
where the eigen states are given by
$|\Gamma_{7\pm} \rangle$=
$\sqrt{1/6}|\pm 5/2\rangle$$-$$\sqrt{5/6}|\mp 3/2\rangle$,
$|\Gamma_{8\pm}^{(1)} \rangle$=
$\sqrt{5/6}|\pm 5/2\rangle$+$\sqrt{1/6}|\mp 3/2\rangle$, 
and
$|\Gamma_{8\pm}^{(2)}\rangle$=$|\pm 1/2\rangle$.
Here $+$ and $-$ in the subscripts denote
``pseudo-spin'' up and down, respectively, within each 
Kramers doublet.



Now we discuss the relative positions of the energy levels
of $\Gamma_{7}$ and $\Gamma_{8}^{(\tau)}$
while taking account of certain features of CeTIn$_{5}$.
Since CeTIn$_5$ has a tetragonal crystal structure
and a quasi two-dimensional Fermi surface \cite{Haga},
it is natural to consider a two-dimensional square lattice
composed of Ce ions.
Due to the effect of anions surrounding the Ce ion, it is deduced
that the energy level of $\Gamma_7$ becomes higher than those of
$\Gamma_8$'s.
Thus, in the following, the $\Gamma_7$ orbital 
is neglected for simplicity.
Note also that we need to include the effect of the tetragonal
CEF, which lifts the degeneracy of the $\Gamma_{8}^{(\tau)}$.
Although a mixing between $\Gamma_{7}$ and $\Gamma_{8}^{(1)}$
generally occurs under a tetragonal CEF,
such a mixing is expected to be small 
since the $\Gamma_7$ orbital has higher energy.
Thus, in this situation, the effect of a tetragonal CEF can be
included in terms of a level splitting $\varepsilon$
between the $\Gamma_{8}^{(1)}$ and $\Gamma_{8}^{(2)}$ orbitals.
In order to explain the magnetic anisotropy
observed in experiments,
$\Gamma_{8}^{(1)}$ must be lower than $\Gamma_{8}^{(2)}$, i.e., 
$\varepsilon$$>$0 in CeTIn$_{5}$.
The magnitude of $\varepsilon$ for each CeTIn$_{5}$ compound
will be discussed later.
It should be noted that the above level scheme is consistent 
with the following two facts:
(i) Experimental results for CeTIn$_{5}$ exhibit a larger uniform
susceptibility for magnetic field perpendicular 
to the CeIn$_{3}$ plane than 
for the parallel case \cite{Ce115,Takeuchi}.
This significant anisotropy is well explained under the assumption 
that $\Gamma_7$ is $not$ the lowest-energy state
and $\varepsilon$ is $positive$.
(ii) The band-structure calculation results suggest that
the almost flat band corresponding to $\Gamma_7$ appears
above the Fermi level \cite{Maehira}. 


In order to include the itinerant features of the $4f$-electrons,
a simple way is to include nearest-neighbor hopping
for the $f$-electrons by the tight-binding method \cite{Maehira,TB}.
Although the hybridization with the In $5p$ electronic states 
may be important,
here such an effect is considered as renormalization
of the effective hopping amplitude of $f$-quasiparticles.
Further, by adding the on-site Coulomb interaction terms
among the $f$-electons, the Hamiltonian becomes
\begin{eqnarray}
  H &=& \sum_{{\bf ia}\tau\tau'\sigma}
  t^{\bf a}_{\tau\tau'} f_{{\bf i}\tau\sigma}^{\dag}
                        f_{{\bf i+a}\tau'\sigma}
  -\varepsilon \sum_{\bf i}(n_{{\bf i}1\sigma}-n_{{\bf i}2\sigma})/2 
  \nonumber \\
  &+&U \sum_{{\bf i}\tau}n_{{\bf i}\tau\uparrow}
                         n_{{\bf i}\tau\downarrow}
  +U'\sum_{{\bf i}\sigma\sigma'}n_{{\bf i}1\sigma}
                                n_{{\bf i}2\sigma'},
\end{eqnarray}
where $f_{{\bf i}\tau\sigma}$ is the annihilation operator 
for an $f$-electron with
pseudo-spin $\sigma$ in the orbital $\Gamma_{8}^{(\tau)}$
at site ${\bf i}$,
${\bf a}$ is the vector connecting nearest-neighbor sites, and
$n_{{\bf i}\tau\sigma}$=$f_{{\bf i}\tau\sigma}^{\dag}
f_{{\bf i}\tau\sigma}$.
The first term represents the nearest-neighbor hopping of $f$-electrons
with the amplitude $t^{\bf a}_{\tau\tau'}$ between
$\Gamma_{8}^{(\tau)}$ and $\Gamma_{8}^{(\tau')}$
along the ${\bf a}$-direction, given by
$t^{\bf x}_{11}$=$-\sqrt{3}t^{\bf x}_{12}$=
$-\sqrt{3}t^{\bf x}_{21}$=$3t^{\bf x}_{22}$=1 
for ${\bf a}$=${\bf x}$ and
$t^{\bf y}_{11}$=$\sqrt{3}t^{\bf y}_{12}$=
$\sqrt{3}t^{\bf y}_{21}$=$3t^{\bf y}_{22}$=1
for ${\bf a}$=${\bf y}$, respectively,
in energy units where $t^{\bf x}_{11}$=1.
Note the positive sign of the first term in $H$,
since the M-point, not the $\Gamma$-point, 
is at the bottom of the bands
forming the Fermi surfaces \cite{Maehira}.
Note also that the present $t^{\bf a}_{\tau\tau'}$ is
just the same as that of the $e_{\rm g}$ electrons
\cite{Dagotto}, but this point will be discussed elsewhere.
The second term denotes the tetragonal CEF.
In the third and fourth terms, 
$U$ and $U'$ are the intra- and inter-orbital Coulomb interations,
respectively.
In reality, $U$=$U'$, since they originate from
the same Coulomb interactions
among $f$-orbitals in the $j$=5/2 multiplet,
but in this paper, we also treat the case
where $U$$\ne$$U'$ in order to analyze the roles of SF and OF.
Since we consider quarter-filling (one $f$-elecron per site),
the model in the limit of $\varepsilon$=$\infty$ reduce
to the half-filled, single-orbital Hubbard model.


Now, in order to investigate superconductivity around 
the spin and/or orbital ordered phases, we calculate 
spin and orbital susceptibilities, 
$\hat{\chi}^{\rm s}({\bf q})$ and $\hat{\chi}^{\rm o}({\bf q})$, 
respectively.
Within the RPA, these are given in a matrix form as
\begin{eqnarray}
  && \hat{\chi}^{\rm s}({\bf q}) \!=\!
  [\hat{1}-\hat{U}^{\rm s}\hat{\chi}({\bf q})]^{-1}
  \hat{\chi}({\bf q}), \\
  && \hat{\chi}^{\rm o}({\bf q}) \!=\!
  [\hat{1}+\hat{U}^{\rm o}\hat{\chi}({\bf q})]^{-1}
  \hat{\chi}({\bf q}),
\end{eqnarray}
where labels of row and column in the matrix appear in the order
11, 22, 12, and 21,
these being pairs of orbital indices 1 and 2.
Note that $\hat{1}$ is the 4$\times$4 unit matrix.
$\hat{U}^{\rm s}$ is given by
$U_{11,11}^{\rm s}$=$U_{22,22}^{\rm s}$=$U$,
$U_{12,12}^{\rm s}$=$U_{21,21}^{\rm s}$=$U'$,
otherwise zero,
while $\hat{U}^{\rm o}$ is expressed as
$U_{11,11}^{\rm o}$=$U_{22,22}^{\rm o}$=$U$,
$U_{11,22}^{\rm o}$=$U_{22,11}^{\rm o}$=$2U'$,
$U_{12,12}^{\rm o}$=$U_{21,21}^{\rm o}$=$-U'$,
otherwise zero.
The matrix elements of $\hat{\chi}({\bf q})$ are defined by
$\chi_{\mu\nu,\alpha\beta}({\bf q})$=
$-T\sum_{{\bf k},n}
G^{(0)}_{\alpha\mu}({\bf k}+{\bf q},i\omega_{n})
G^{(0)}_{\nu\beta}({\bf k},i\omega_{n})$,
where $T$ is temperature,
$G^{(0)}_{\mu\nu}({\bf k},i\omega_{n})$
is the non-interacting Green's function for $f$-electrons
with momentum ${\bf k}$ propagating between $\mu$- and $\nu$-orbitals,
and $\omega_n$=$\pi T(2n$$+$$1)$ with integer $n$.
The instabilities for the spin- and orbital-ordered phases
are determined by the conditions of
${\rm det}[\hat{1}-\hat{U}^{\rm s}\hat{\chi}({\bf q})]$=0
and ${\rm det}[\hat{1}+\hat{U}^{\rm o}\hat{\chi}({\bf q})]$=0, 
respectively.

By using $\hat{\chi}^{\rm s}({\bf q})$ and 
$\hat{\chi}^{\rm o}({\bf q})$,
the superconducting gap equation is given as
\begin{equation}
  \label{gap}
  {\bf \Delta}^{\xi}({\bf k})=\sum_{{\bf k'}}
    \hat{V}^{\xi}({\bf k}-{\bf k'})
    \hat{\phi}({\bf k'}){\bf \Delta}^{\xi}({\bf k'}),
\end{equation}
where ${\bf \Delta}^{\xi}({\bf k})$
=[$\Delta^{\xi}_{11}({\bf k})$, $\Delta^{\xi}_{22}({\bf k})$,
$\Delta^{\xi}_{12}({\bf k})$, $\Delta^{\xi}_{21}({\bf k}$)]$^t$
is the gap function in the vector representation
for a singlet ($\xi={\rm S}$) or triplet 
($\xi={\rm T}$) pairing state,
and the matrix elements of the singlet- and triplet-pairing 
potentials, respectively, are given by
\begin{eqnarray}
  &&V^{\rm S}_{\alpha\beta,\mu\nu}({\bf q})=
    [(-3/2)\hat{W}^{\rm s}({\bf q})+(1/2)\hat{W}^{\rm o}({\bf q})
    +\hat{U}^{\rm s}]_{\alpha\mu,\nu\beta},\\
  &&V^{\rm T}_{\alpha\beta,\mu\nu}({\bf q})=
    [(1/2)\hat{W}^{\rm s}({\bf q})+(1/2)\hat{W}^{\rm o}({\bf q})
    -\hat{U}^{\rm s}]_{\alpha\mu,\nu\beta}.
\end{eqnarray}
Here, the spin and orbital susceptibilities are included as
$\hat{W}^{\rm s}({\bf q})$=$\hat{U}^{\rm s}
+\hat{U}^{\rm s}\hat{\chi}^{\rm s}({\bf q})\hat{U}^{\rm s}$ and 
$\hat{W}^{\rm o}({\bf q})$=$-\hat{U}^{\rm o}
+\hat{U}^{\rm o}\hat{\chi}^{\rm o}({\bf q})\hat{U}^{\rm o}$.
The elements of the pair correlation function
$\hat{\phi}({\bf k})$ are given by
$\phi_{\alpha\beta,\mu\nu}({\bf k})$=$
T\sum_{n}
G^{(0)}_{\alpha\mu}({\bf k},-i\omega_{n})
G^{(0)}_{\nu\beta}({\bf -k},i\omega_{n})$.
The superconducting transition is obtained for each respective 
irreducible representation by solving Eq.~(\ref{gap}), 
where the maximum eigenvalue becomes unity.


Here we mention the essential symmetries of the
Cooper pairs in multi-orbital systems.
The pairing states are classified into four types
owing to spin $SU(2)$ symmetry and space inversion symmetry;
(1) spin-singlet and orbital-symmetric with even-parity,
(2) spin-triplet and orbital-symmetric with odd-parity,
(3) spin-singlet and orbital-antisymmetric with odd-parity,
and 
(4) spin-triplet and orbital-antisymmetric with even-parity.
Note that $SU(2)$ symmetry does not exist in orbital space,
since the lattice and the local wavefunctions rotate simultaneously.
For (1) and (2), $f$-electrons in the same band
form the Cooper pair, while for (3) and (4),
such a pair is formed only between different bands.
Except for the special case in which two Fermi surfaces
connect with each other,
orbital-antisymmetric pairing is unstable due to
depairing effects destroy inter-band pairs.
In fact, even after careful calculations,
we do not find any region for (3) and (4) in the parameter space 
considered here.
Thus, in the following, we discuss only the orbital-symmetric pair.

\begin{figure}[t]
\centerline{\epsfxsize=8.5truecm \epsfbox{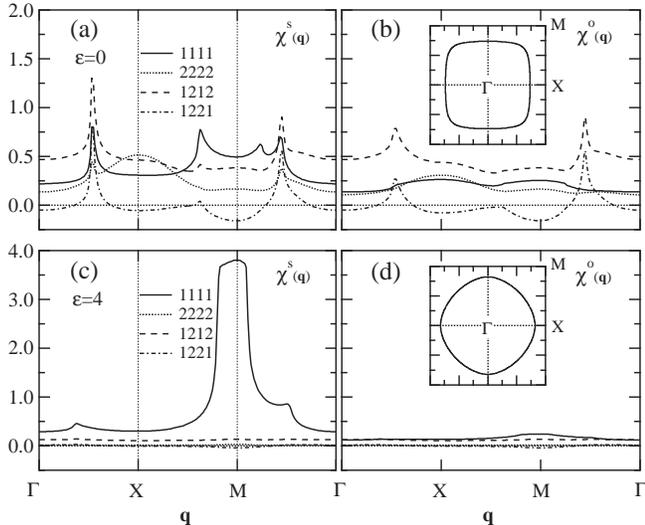} }
\caption{(a) Spin  and (b) orbital susceptibilities in {\bf q} space
for $\varepsilon$=0 and $U$=$U'$=0.9$U_{\rm m}$. 
(c) and (d) are for $\varepsilon$=4.
Insets indicate the Fermi-surface lines.}
\end{figure}

Let us discuss first the effect of $\varepsilon$ on the correlation
between SF and OF.
Since we are considering a realistic situation 
corresponding to CeTIn$_5$,
we restrict ourselves to the case of $U$=$U'$ for a while.
In Fig.~1, the principal components of 
$\hat{\chi}^{\rm s}({\bf q})$ and
$\hat{\chi}^{\rm o}({\bf q})$ are shown in ${\bf q}$ space
for $U$=$U'$=0.9$U_{\rm m}$,
where $U_{\rm m}$ denotes spin instability determined from
${\rm det}[\hat{1}-\hat{U}^{\rm s}\hat{\chi}({\bf q})]$=0.
In the orbital-degenerate case with $\varepsilon$=0,
as shown in Figs.~1(a) and (b),
the overall magnitude of OF is comparable with that of SF.
As for the ${\bf q}$ dependence, common structures are found in
$\chi^{\rm s}({\bf q})$ and $\chi^{\rm o}({\bf q})$.
Namely, there are two peaks around $(\pi/2,\pi/2)$ and $(\pi/2,0)$ 
owing to the nesting properties of the Fermi surface (see the inset).
Thus, the orbital-degenerate region is characterzied 
by competition between SF and OF.
With increasing $\varepsilon$, 
the Fermi surface approaches a shape
having the nesting vector $(\pi,\pi)$ (see the inset).
For $\varepsilon$=4, as shown in Fig.~1(c) and (d), 
OF is almost completely suppressed,
and SF around $(\pi,\pi)$ become dominant.
Note here that the increase of $\varepsilon$ 
makes the lower energy state
favorable and suppresses excitations to the upper energy state,
indicating the suppression of OF.
Thus, even at this stage, it is understood that
the suppression of OF leads to the development of SF for
the paramagnetic system which is wavering 
between spin- and orbital-ordering.


Next, we consider how a superconducting phase emerges
when $\varepsilon$ is increased.
In Fig.~2(a), the maximum eigenvalues 
for several irreducible representations
are depicted as a function of $\varepsilon$ for $U$=$U'$=2.5.
The calculations are carried out for a fixed $T$=0.02,
and the first Brillouin zone is divided into $128\times128$ meshes.
In the orbital-degenerate region, several eigenvalues are
very close to each other owing to multi-peak structures
in $\hat{\chi}^{\rm s}({\bf q})$ and $\hat{\chi}^{\rm o}({\bf q})$.
With increasing $\varepsilon$, 
as is easily understood from the growth of AFSF mentioned above,
the eigenvalue for $B_{\rm 1g}$ symmetry becomes dominant
and finally at $\varepsilon$$\approx$3, 
the $B_{\rm 1g}$ superconducting phase
is stabilized.
On further increasing $\varepsilon$, 
the AF instability eventually occurs,
since the system asymptotically
approaches the half-filled single-orbital Hubbard model.
We emphasize that the increase of $\varepsilon$ brings transitions
successively
in the order of PM, $B_{\rm 1g}$ superconducting, and AF phases.

\begin{figure}[t]
\centerline{\epsfxsize=8.5truecm \epsfbox{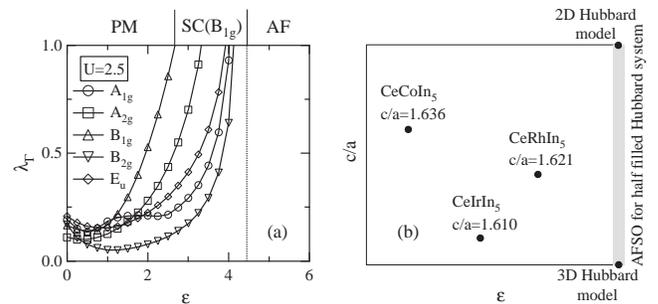} }
\caption{
(a) Maximum eigenvalue vs. $\varepsilon$ for the respective irreducible
representation at $U$=$U'$=2.5. 
(b) Schematic plot $c/a$ vs. $\varepsilon$ to illustrate a comparison
between our theory and actual CeTIn$_5$ compounds.}
\end{figure}

Based on the above calculated results,
let us try to explain the differences among three Ce-115 compounds,
CeRhIn$_{5}$ (N\'eel temperature $T_{\rm N}$=3.8K),
CeIrIn$_{5}$ (superconducting transition temperature $T_{\rm c}$=0.4K),
and CeCoIn$_{5}$ ($T_{\rm c}$=2.3K) \cite{Ce115}.
One property which distinguishes these compounds is
two-dimensionality,
as expressed by the ratio $c/a$ between lattice constants.
It is naturally expected that two-dimensionality becomes stronger
in the order of 
CeIrIn$_{5}$, CeRhIn$_{5}$, and CeCoIn$_{5}$ \cite{Ce115}.
Since the magnetic state should be stabilized with increasing 
three-dimensionality,
CeIrIn$_{5}$ should be most favorable for the occurence of 
antiferromagnetism among three compounds,
but this is obviously inconsistent with experimental results.
This inconsistency is resolved by introducing
another important ingredient $\varepsilon$.
One can show that the increase of magnetic anisotropy
just corresponds to the increase of $\varepsilon$.
Analysis of experimental results for the anisotropy of
magnetic susceptibilities \cite{Ce115,Takeuchi} leads
to the conclusion  
that $\varepsilon$ becomes larger 
in the order of CeCoIn$_{5}$, CeIrIn$_{5}$, and CeRhIn$_{5}$. 
Taking into account the effect of $\varepsilon$ in addition to 
dimensionality, we arrive at the picture  
schematically shown in Fig. 2(b).
Due to the enhancement of AFSF induced by 
increasing $\varepsilon$, as shown in Fig.~1,
it may be understood that CeRhIn$_{5}$ rather than CeIrIn$_{5}$
is more favorable to antiferromagnetism. 
Namely, it is considered that 
CeRhIn$_{5}$ with the largest $\varepsilon$ is antiferromagnet,
while CeCoIn$_{5}$ and CeIrIn$_{5}$ with smaller $\varepsilon$ 
than CeRhIn$_{5}$ exhibit superconductivity.
The difference in $T_{\rm c}$ between 
CeCoIn$_{5}$ and CeIrIn$_{5}$ may be attributed to
the extent of two-dimensionality. 
Concerning the superconductivity and antiferromagnetism 
of CeTIn$_{5}$ compounds,
the combination of 
two dimensionality and the crystalline field splitting
is necessary to reach a consistent picture.

\begin{figure}[t]
\centerline{\epsfxsize=8.5truecm \epsfbox{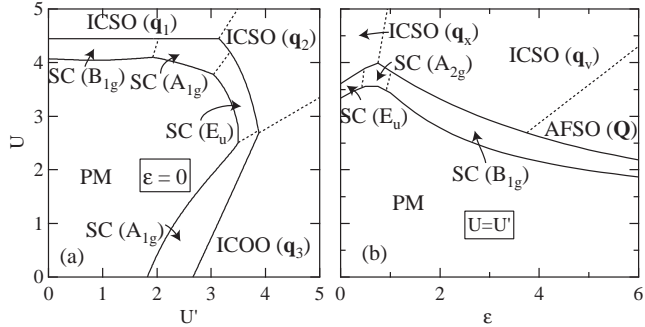} }
\caption{Phase diagram (a) in the $U$-$U'$ plane for $\varepsilon$=0
and (b) in the $U$-$\varepsilon$ plane for $U$=$U'$.
Meanings of abbreviations used here are as follows:
SC($\Gamma$) denotes superconducting phase with $\Gamma$-symmetry,
ICSO($\bf q$) indicates incommensurate spin ordered phase with
wave vector ${\bf q}$,
ICOO($\bf q$) is incommensurate orbital ordered phase
with wave vector ${\bf q}$,
and AFSO($\bf Q$) denotes antiferromagnetic spin ordered phase. 
Several wave vectors are defined as
${\bf q}_{1}$=$(0.56\pi,0.56\pi)$,
${\bf q}_{2}$=$(0.53\pi,0)$,
${\bf q}_{3}$=$(\pi,0)$,
${\bf q}_{x}$=$(q_{x},0)$,
${\bf q}_{\rm v}$=$(\pi,q_{y})$,
and ${\bf Q}$=$(\pi,\pi)$.
Solid curves are the phase boundaries 
determined by actual calculations,
while dotted lines are schematic phase boundary guides for the eye.}
\end{figure}

The orbitally degenerate model, Eq.~(1), shows new, rich 
superconducting properties both for singlet and triplet pairings. 
In Fig.~3(a) and (b), we show the phase diagrams 
in the $U$-$U'$ plane for
$\varepsilon$=0 and $U$-$\varepsilon$ plane for $U$=$U'$, 
respectively. 
In these figures, 
four characteristic superconducting phases are observed: 
(1) a SF-mediated spin-singlet superconducting phase 
with $B_{\rm 1g}$-symmetry in the vicinity of a spin-ordered phase;
(2) a $g$-wave superconducting phase with $A_{\rm 2g}$-symmetry;
(3) a spin-triplet superconducting phase with $E_{\rm u}$-symmetry 
due to the cooperation between SF and OF 
around $U$=$U'$ and $\varepsilon$=0 and
(4) an OF-mediated spin-singlet superconducting phase
with $A_{\rm 1g}$-symmetry for $U$$\ll$$U'$ 
around an orbital-ordered phase.
With respect to (1), it is thought that 
$B_{\rm 1g}$-superconductivity is induced by SF around 
$\bf q$=$\bf Q$=$(\pi,\pi)$. 
Concerning the triplet superconductivity (3), 
note that for this region all superconducting instabilities are 
quite close to each other because of the multi-peak structures in 
$\hat{\chi}^{\rm s}({\bf q})$ and $\hat{\chi}^{\rm o}({\bf q})$ 
as shown in Fig. 1.
We also note that the factors in front of
the SF term in $\hat{W}^{\rm s}({\bf q})$ are $-3/2$ for singlet and
$+1/2$ for triplet pairing, while the factor for the OF term
in $\hat{W}^{\rm o}({\bf q})$ is $+1/2$ for both pairings.
Namely, OF is cooperative with SF for spin-triplet pairing,
while SF and OF compete with each other for spin-singlet pairing
\cite{Takimoto}.
Thus, the spin-triplet pairing phase appears 
in the region for $\varepsilon$$\approx$0 and $U$$\approx$$U'$.
The mechanism of the OF-mediated singlet supercondutivity (4) 
is interesting.
In this state, two quasi-particles with different pseudo-spin 
form an on-site intra-orbital singlet pair since $U'{\gg}U$, 
leading to $A_{\rm 1g}$ superconductivity.
In short, singlet superconductivity with $B_{\rm 1g}$- 
or $A_{\rm 1g}$-symmetry is stabilized 
when either SF or OF is dominant, 
while triplet superconductivity is favored 
when there is cooperation between SF and OF.


In summary, we have studied superconductivity in the Ce-115 systems
based on the orbitally degenerate Hubbard model,
and found that with increasing the tetragonal CEF splitting 
$d_{x^{2}-y^{2}}$-wave superconductivity due to AFSF
emerges out of the suppression of OF in the vicinity of the
AF phase.
The concept of AFSF-induced superconductivity controlled by OF
qualitatively explains the difference among Ce-115 materials.


The authors thank H. Harima, A. Hasegawa, T. Moriya,
Y. \=Onuki, R. E. Walstedt, and K. Yamada for discussions.
K.U. is financially supported by
Grant-in-Aid for Scientific Research Areas (B)
from the Ministry of Education, Science, Sports, Culture, and
Technology.

\vskip-0.5cm

\end{document}